\documentclass[conference]{IEEEtran}
\IEEEoverridecommandlockouts
\usepackage{cite}
\usepackage{amsmath,amssymb,amsfonts}
\usepackage{algorithmic}
\usepackage{graphicx}
\usepackage{textcomp}
\usepackage{xcolor}

\def\bk{{\boldsymbol k}}
\def\bh{{\boldsymbol h}}

\def\bC{{\boldsymbol C}}           
 
\def\bI{{\boldsymbol I}}
\def\b0{{\boldsymbol 0}}
\def\bM{{\boldsymbol M}}

\def\be{{\boldsymbol e}}
\def\bh{{\boldsymbol h}}

\def\balpha{{\boldsymbol \alpha}}

\def\BibTeX{{\rm B\kern-.05em{\sc i\kern-.025em b}\kern-.08em
    T\kern-.1667em\lower.7ex\hbox{E}\kern-.125emX}}
\begin{document}

\title{Mathematical modelling of nuclear medicine data
}

\author{\IEEEauthorblockN{1\textsuperscript{st} Michele Piana}
\IEEEauthorblockA{\textit{dipartimento di matematica} \\
\textit{universita di genova}\\
Genova, Italy \\
piana@dima.unige.it}
\and
\IEEEauthorblockN{2\textsuperscript{nd} Giacomo Caviglia}
\IEEEauthorblockA{\textit{dipartimento di matematica} \\
\textit{universita di genova}\\
Genova, Italy \\
caviglia@dima.unige.it}
\and
\IEEEauthorblockN{3\textsuperscript{rd} Sara Sommariva}
\IEEEauthorblockA{\textit{dipartimento di matematica} \\
\textit{universita di genova}\\
Genova, Italy \\
sommariva@dima.unige.it}
}

\maketitle

\begin{abstract}
Positron Emission Tomography using 2-[18F]-2deoxy-D-glucose as radiotracer (FDG-PET) is currently one of the most frequently applied functional imaging methods in clinical applications. The interpretation of FDG-PET data requires sophisticated mathematical approaches able to exploit the dynamical information contained in this kind of data. Most of these approaches are formulated within the framework of compartmental analysis, which connects the experimental nuclear data with unknown tracer coefficients measuring the effectiveness of the tracer metabolism by means of Cauchy systems of ordinary differential equations. This paper provides a coincise overview of linear compartmental methods, focusing on the analytical solution of the forward compartmental problem and on the specific issues concerning the corresponding compartmental inverse problem. 
\end{abstract}

\begin{IEEEkeywords}
component, formatting, style, styling, insert
\end{IEEEkeywords}

\section{Introduction}
 The metabolic pattern of most solid tumors shows an increased glucose consumption, even under aerobic conditions \cite{VanderHeiden}. The mechanisms underlying this effect are not completely known, but a number of studies documented a direct relationship between glucose consumption and aggressiveness  in cancer tissues. Although direct measurement of the continuous flux of glucose molecules through lesion-populating cells is extremely difficult, a reliable estimate was made possible by the peculiar kinetic features of the radioactive glucose analogue 2-[18F]-2deoxy-D-glucose (FDG).
 
In vivo, cancer FDG retention is dependent upon blood glucose level {\cite{Williams,SciRep}, drugs \cite{Garbarino_kidney}, and overall tracer availability in blood, which in turn depends on the amount of administered activity and diffusion throughout the whole body, after injection. Tracer concentration in blood also varies with time as a consequence of physiological factors, related to urinary elimination  \cite{Garbarino_kidney}, accumulation in liver  \cite{Garbarino_liver}, absorption by brain, and the different accumulation rates of the various tissues \cite{Busing}. 

Positron emission tomography (PET) measures the radiation emitted by the target tissue in vivo, following an intravenous administration of tracer molecules. The measuring device is calibrated so that the activity distribution inside the tissue is reconstructed. The output may vary from the estimate of the concentration of activity in a subregion of the tissue at a chosen instant, to the time course of the activity in a given time interval. The independent time variable $t$ [min] measures the time interval from tracer infusion. In particular, FDG can be used as injected tracer and FDG-PET is currently the most used modality in this kind of functional imaging analysis. 

In the present paper we provide an overview of how mathematical modelling can be used to obtain a reliable interpretation of FDG-PET data. Specifically, we will show how systems of ordinary differential equations can be used to connect the experimental data provided by FDG-PET to unknown kinetic parameters that describe the effectiveness of FDG metabolism in biological tissue.

The plan of the paper is as follows. Section 2 will introduce compartmental analysis, as the most reliable tool for the mathematical modelization of FDG-PET data. Section 3 will provide some results concerning the forward problem associated to compartmental analysis and Section 4 will briefly illustrate some aspects of the corresponding inverse problem. Our conclusions will be offered in Section 5.

\section{Compartmental analysis}

Compartmental analysis \cite{Logan,Patlak} provides a mathematical model relating PET image data to specific metabolic states or chemical compounds of the tracer, taking also into account their distribution in space, if needed. The metabolic states are known as comportments, sources, or pools.  A fundamental requirement of compartmental modelling is the so-called well-mixed assumption, which means that the tracer distribution in each compartment is spatially homogeneous, and tracer exchanged between compartments in instantaneously mixed. Further conditions for applicability of compartmental analysis can be found in \cite{Bertoldo, Cherry, Muzi, Wernick}.  

Each compartment denotes a specific metabolic condition of FDG in the biological tissue. For example, FDG may be free in the interstitial space between cellular membranes or trapped within the cytosol by means of a biochemical process named phosphorylation. Further, each compartment is characterized by the related time dependent activity concentration. Finally, it is understood that different compartments, such as free and phosphorylated FDG, may occupy the same spatial volume; conversely, if a chemical compound of the tracer occupies volumes separated by a membrane, as occurs to free tracer in interstitial tissue and cytosol, then the compound is associated with two spatially distinct compartments, of possibly different concentrations.   

A compartmental model (CM) is given by an interconnected set of compartments. The number of compartments to be considered depends on the chemical, physiological, and biological properties of the tracer to be modelled \cite{Watabe, Cherry,Lawson}. We point out that that, in principle, blood should be considered as a compartment; however, in compartmental analysis the concentration of tracer inside blood is always considered as known via either experimental measurements or computational approaches. This specific concentration is called input function (IF) and in the following it will be considered as one of the problem data \cite{Wernick}. Therefore, the main ingredients of a compartmental model for FDG-PET data relies on the following items:
\begin{itemize}
\item The concentrations of the various pools are the state variables of the CM.
\item The time dependence of the state variables is determined by tracer exchange. 
\item The flux of tracer between compartments occurs according to mass conservation. 
\item The time rate of the concentration of each compartment is set equal to the difference between the tracer that enters and leaves the compartment, per unit time and unit volume. 
\item The IF is the forcing function of the system providing tracer supply to the interconnected compartments.
\end{itemize}
Application of the conservation law shows that concentrations are related by a system of ordinary differential equations (ODEs).  Here it is assumed that all the initial concentrations vanish, because there is no tracer available at the beginning of each experiment. According to this mathematical model, the state variables are the solutions of a Cauchy problem. We consider linear ODEs with constant rate coefficients, representing the rate of flux of tracer between compartments (for example, the rate of phosphorylation of FDG molecules). The constants are also termed  transfer coefficients or microparameters \cite{Gunn}. If the rate coefficients and the initial state are given, then the solution of the Cauchy problem (i.e., the compartmental forward problem) provides a detailed description of tracer kinetics. However, typical problems of compartmental analysis require the determination of the rate constants such that the corresponding solution complies with the measured overall tissue concentration. From the viewpoint of mathematics, this is a typical inverse problem.  

\section{Compartmental analysis: forward problem}
The mathematical complexity of the forward problem for compartmental analysis depends on the number of compartments considered. As a starting point, the 1-compartment model (1-CM) is an oversimplified model where there is only one tissue compartment $\mathcal{C}_f$ with state variable $C_f$, accounting for the overall tracer content, which is considered as free (i.e., not phosphorylated). The differential equation for $C_f$ is
\begin{equation} \label{eq:EDO_1C}
\dot{C}_f = -k_2 \,C_f + k_1 \, C_b ~,
\end{equation}
where $C_b$ is the IF (i.e., the concentration in the blood compartment), $k_1$ and $k_2$ are the rate constants for incoming and outgoing tracer. In other words, $k_1 \,C_b$ is the rate of incoming flow of tracer per unit volume, while $k_2\,C_f$ is the outgoing flow per unit volume; thus, the net rate of tracer concentration per unit volume, $\dot{C}_f$, is the difference between incoming and outgoing flows, consistently with conservation of the mass of tracer (note that in CM equations, the plus and minus signs refer systematically to incoming and outgoing flows, respectively, of the compartment considered). The solution of equation (\ref{eq:EDO_1C})  (vanishing at $t=0$) is
\begin{equation} \label{eq:sol_1C}
C_f =  k_1 \, \int_0^t  e^{- k_2 \,(t-\tau) }  \, C_b (\tau) \,d\tau. 
\end{equation}
The concentration $C_f$ is proportional to $k_1$, which is related to the absorption capacity of the tissue. The parameter $1/k_2$ is related to the asymptotic equilibrium time. 
 
Standard 2-compartment models (2-CMs) have been developed under the assumption that the intracellular processes of phosphorylation and dephosphorylation of FDG are modelled by the use of two compartments  $\mathcal{C}_f$ and $\mathcal{C}_p$, accounting for free and phosphorylated tracer, respectively (see, e.g., \cite{Sokoloff, Wernick} ). However, more recent advances in biochemistry show that G6Pase is anchored to the endoplasmic reticulum (ER) \cite{Ghosh}, so that its action of hydrolysis of G6P and FDG6P, resulting in the creation of a phosphate group and free molecules of glucose and FDG, occurs after transport of the phosphorylated forms into the ER by the transmembrane protein glucose-6-phosphate transporter (G6PT) \cite{Marini}. 
Then, free FDG in ER may be  released in cytosol. Further biochemical, pharmacological, clinical, and genetic data lead to a natural interpretation of the ER as a distinct metabolic compartment \cite{Csala} and therefore to the formulation of a 3-compartment model (3-CM). 

From a formal viewpoint, 2-CM and 3-CM can be described in a matrix form as follows:

\begin{equation} \label{eq:EDO_comp}
\dot{\bC} = \bM \, \bC + k_1 \, C_b \, \be ~.
\end{equation}
In general, $\bC$ is the n-dimensional column vector of state variables, with $n=2,3$; $\bM$ is a constant square matrix of order $n$, with entries given by the rate coefficients; $\be$ is a constant $n$-dimensional column vector. Addition of the initial condition $\bC(0)=0$ gives rise to a Cauchy problem for the unknown state vector $\bC$. For 2-CMs we have that
\begin{equation} \label{eq:def_A_C_e_2-C}
 \bM = \begin{bmatrix}  -(k_2+k_3) & k_4 \\ k_3 & -k_4 \end{bmatrix}, \quad
\bC = \begin{bmatrix}  C_f  \\ C_p \end{bmatrix}, \quad \be = \begin{bmatrix}  1  \\ 0 \end{bmatrix} \end{equation}
while 3-CM equations can be written as
\begin{equation} \label{eq:def_A_C_e_3-C}
 \bM = \begin{bmatrix}  -(k_2+k_3) & 0 & k_6 \\ k_3 & -k_5 & 0 \\ 0 & k_5 & - k_6 \end{bmatrix}, \quad
\bC = \begin{bmatrix}  C_f  \\ C_p  \\ C_r \end{bmatrix}, \quad \be = \begin{bmatrix}  1  \\ 0 \\ 0 \end{bmatrix}. \end{equation}

The solution of the direct problem may be represented as
\begin{equation} \label{eq:sol_comp}
\bC(t; \bk,C_b) = k_1 \, \int_0^t e^{\bM \,(t-\tau)} \,  \, \, C_b(\tau) \, d\tau   \, \be.
\end{equation}
where $\bk$ is the vector of parameters defined as $\bk= (k_1, k_2, \dots , k_m)^T$, with upper $T$ denoting the transpose and $m$ depending on the CM adopted. The notation gives evidence to the dependence of $\bC$ on the rate constants and the input function. \\

\section{Compartmental analysis: inverse problem}
The activity concentration measured by a PET scanner results from superposition of various signals emitted, e.g., by tracer molecules carried by blood partially occupying the ROI volume, molecules dispersed in the interstitial tissue, free and phosphorylated molecules in cells. The tissue selected for measurement of the activity concentration $C_T$  [Bq/ml] is referred to as the target tissue. The time course of $C_T$, also regarded as the tissue response, is obtained from the ROI analysis of a dynamic series of images (see, e.g., \cite{Cherry, Wernick}). We recall that data are corrected for attenuation and, possibly, other systematic sources of error. Obviously, the reconstructed value of $C_T$ is functionally dependent on characteristics of the IF and of the target tissue, the injected dose, and the physiologic conditions of the patient. From a mathematical viewpoint, the connection between the compartmental model and the measured PET data is given by the inverse problem equation
\begin{equation} \label{eq:dato_comp}
C_T = \balpha^T\, \bC(t;\bk,C_b)~, 
\end{equation} 
where $\balpha$ is a row vector with components depending on physiological parameters.  

As a first step in the solution of the inverse problem, it must be shown that the rate coefficients are uniquely determined by data. This requirement puts severe limitations on the number of coefficients allowed, and hence on the number of compartments-metabolic states and the related interconnections to be considered. The result is a sort of compromise between the need for simplification in the formal description and exhaustiveness in the representation of reality. Here we consider models resulting from either two or three compartments. In the case of linear compartmental models as the ones considered in this paper, a standard procedure for proving uniqueness results is possible: denoting as $\tilde{f}(s)$ the Laplace transform of a function $f(t)$ and assuming that appropriate regularity conditions are satisfied, the procedure can be sketched as follows:
\begin{enumerate}
\item Consider the time derivative of both sides of (\ref{eq:dato_comp}) at the time $t=0$:
\[
\dot{C}_T(0) = \balpha\, \dot{\bC}(0)+ V_b \,\dot{C}_i(0). 
\]
According to eq (\ref{eq:EDO_comp}), $\dot{\bC}(0)$ is replaced by  $k_1 \, C_b(0) \, \be$.  
Hence it follows by substitution that
\begin{equation} \label{eq:k_1}
k_1 = \frac{  \dot{C}_T(0) - V_b \,\dot{C}_i(0)}{ \balpha \, \be \, C_b(0)}.
\end{equation}
This shows that, in principle, $k_1$ is directly determined by data, if $C_b(0) >0$.
\item Consider the Laplace transform of system (\ref{eq:EDO_comp}), with vanishing initial conditions, and equation (\ref{eq:dato_comp}). Then
\begin{equation} \label{eq:LT_EDO}
(s\, \bI- \bM )\, \tilde{\bC} = k_1 \, \tilde{C}_i \, \be 
\end{equation}
\begin{equation} \label{eq:LT_dato}
\tilde{C}_T = \balpha\, \tilde{\bC} + V_b \,\tilde{C}_i, 
\end{equation} 
where $\bI$ denotes the identity matrix. 
\item Consider the solution $ \tilde{\bC}$ of the linear system (\ref{eq:LT_EDO}), which is given by
\begin{equation} \label{eq:tilde_C}
\tilde{\bC} = k_1 \, (s\, \bI- \bM )^{-1} \, \be \,\tilde{C}_i  = k_1 \, \frac{Q(s,\hat{\bk})} {D(s;\hat{\bk})}\, \tilde{C}_i 
\end{equation}
where $D(s,\hat{\bk})$ is the determinant of the matrix $s\,\bI-\bM$, expressed as a polynomial of $s$ of degree $n$, with coefficients depending on $\hat{\bk}=[k_2, ..., k_m]^T$; similarly, $Q(s,\hat{\bk})$ is a polynomial of degree smaller than $n$. 
Substitution into (\ref{eq:LT_dato}) of $ \tilde{\bC}$ leads to  
\begin{equation} \label{eq:Q/P}
\tilde{C}_T - V_b \,\tilde{C}_i = k_1 \, \frac{Q(s,\hat{\bk})} {D(s;\hat{\bk})}\, \tilde{C}_i. 
\end{equation} 
where $\tilde{C}_T$ and $\tilde{C}_i$ are given.
\item Any alternative set $\bh$ of kinetic parameters consistent with the data must satisfy equation (\ref{eq:Q/P}), with $\bk$ replaced by $\bh$. It follows by comparison that
\begin{equation} \label{eq:h_e_k}
  k_1 \, \frac{Q(s,\hat{\bk})} {D(s;\hat{\bk})} =  h_1 \, \frac{Q(s,\hat{\bh})} {D(s;\hat{\bh})}.
\end{equation}
We know from (\ref{eq:k_1}) that $h_1=k_1$.
Thus, if there are no common roots between the polynomials $Q$ and $D$, then equation (\ref{eq:h_e_k}) is equivalent to
\begin{equation} \label{eq:h_e_k_fin}
   Q(s,\hat{\bh}) = Q(s;\hat{\bk}) , \qquad D(s,\hat{\bh}) = D(s;\hat{\bk}) .
\end{equation}
\item If equations (\ref{eq:h_e_k_fin}) imply $\hat{\bh}=\hat{\bk}$, then identifiability is proved.\\
\end{enumerate}
We conclude this general discussion on the uniqueness issue with three comments:
\begin{itemize} 
\item In general, point 5) may be extremely challenging and specific mathematical assumptions must be introduced to deal with it. 
\item If physiological parameters contained in $\balpha$ are considered as unknowns in the inverse problem, then they have to be properly considered in the discussion of identifiability. 
\item There are still open questions on the analysis of identifiability for nonlinear compartmental systems, e.g., when fluxes between compartments are modelled by the Michaelis-Menten law; a comparison of currently available techniques can be found in \cite{Chis}.   
\end{itemize}

In conclusion of this Section, we point out that the second aspect concerning the compartmental inverse problem is related to the numerical solution of equation (\ref{eq:dato_comp}). From an inverse problem perspective, that equation can be summarized as
\begin{equation}\label{frechet-1}
C_T = {\cal{F}}_t(\bk)~,
\end{equation}
where ${\cal{F}}_t: \mathbb{R}^n_+ \rightarrow C^1(\mathbb{R}_+,\mathbb{R})$ is given by
\begin{equation}\label{frechet-2}
{\cal{F}}_t(\bk)=\balpha^T\, \bC(t;\bk,C_b)~.
\end{equation}
Equation (\ref{frechet-1}) is clearly a non-linear zero-finding problem that can be numerically addressed by means of optimization algorithms. It must me pointed out that the operator ${\cal{F}}_t$ is compact and therefore problem (\ref{frechet-1}) is ill-posed in the sense of Hadamard \cite{bepi06}. It follows that, at some stage of the optimization process, a regularization step must be introduced \cite{gaetal13,scetal17}.

\section{Conclusions}
This paper provides a coincise overview of some of the mathematical aspects of compartmental analysis. We formulated the problem in the case of linear models and considered both the solution of the Cauchy system in the case of 2D and 3D models, and the discussion of uniqueness issues associated to the inverse problem.

Compartmental analysis is still an important mathematical tool for the interpretation of nuclear medicine data for many different physiological and clinical applications. Current development of this approach include parametric imaging methods that process PET raw data in order to provide spatially-resolved maps of the kinetic parameters, applications to experimental modalities other than FDG-PET (for example, data acquired by means of ligand tracer technologies), and formulation of models that benefit of exploiting more details physiological information within the compartmental equations framework.

%


%
%
%

\vspace{12pt}

\end{document}